\documentclass{mem}
\usepackage{natbib}\usepackage{txfonts}\usepackage{balance}
\usepackage{graphicx}
\usepackage[a4paper]{hyperref}
\idline{75}{282}
\begin{document}
\def\teff{$T\rm_{eff }$}
\def\kms{$\mathrm {km s}^{-1}$}

\title{
Helium diffusion during the evolution of solar-type stars : asteroseismic tests
}

\author{
M. \,Castro 
\and S. \,Vauclair
          }

  \offprints{M. Castro}

\institute{
Laboratoire d'Astrophysique de Toulouse-Tarbes --
Observatoire Midi-Pyr\'en\'ees, 14 av. Edouard Belin,
31400 Toulouse, France
\email{mcastro@ast.obs-mip.fr}
}

\authorrunning{Castro et al.}

\titlerunning{Helium diffusion in solar-type stars}

\abstract{The element diffusion, described by \cite{michaud70}, is now recognized to occur in all kinds of stars. We attempt to give evidence of signatures of helium diffusion below the convective zone by the way of asteroseismology.
\keywords{Stars: abundances --
Stars: oscillations }
}
\maketitle

Under the opposite effects of gravitation and thermal settling on the one hand, and radiative acceleration on the other hand, the various elements undergo relative separation. In solar-type stars, the outer convective layer let helium falling down during the evolution on the main sequence, enhancing the helium gradient below the convective layer.
\\

We compute models, with masses from 1.1 $M_{\odot}$ to 1.5 $M_{\odot}$, using the Toulouse-Geneva evolution code. We compare couples of models, observationally identical, i.e. with the same external parameters (\teff, L, chemical composition), but in each couple element diffusion is introduced in one of the models, not in the other. This calibration is achieved by adjusting the two free parameters of the stellar evolution code: the initial helium abundance $Y_{0}$ and the mixing length parameter $\alpha$. For each couple, the two models lie at the same point in the HR diagram.

\begin{figure}
 \centering
 \resizebox{6.5cm}{6cm}{\includegraphics{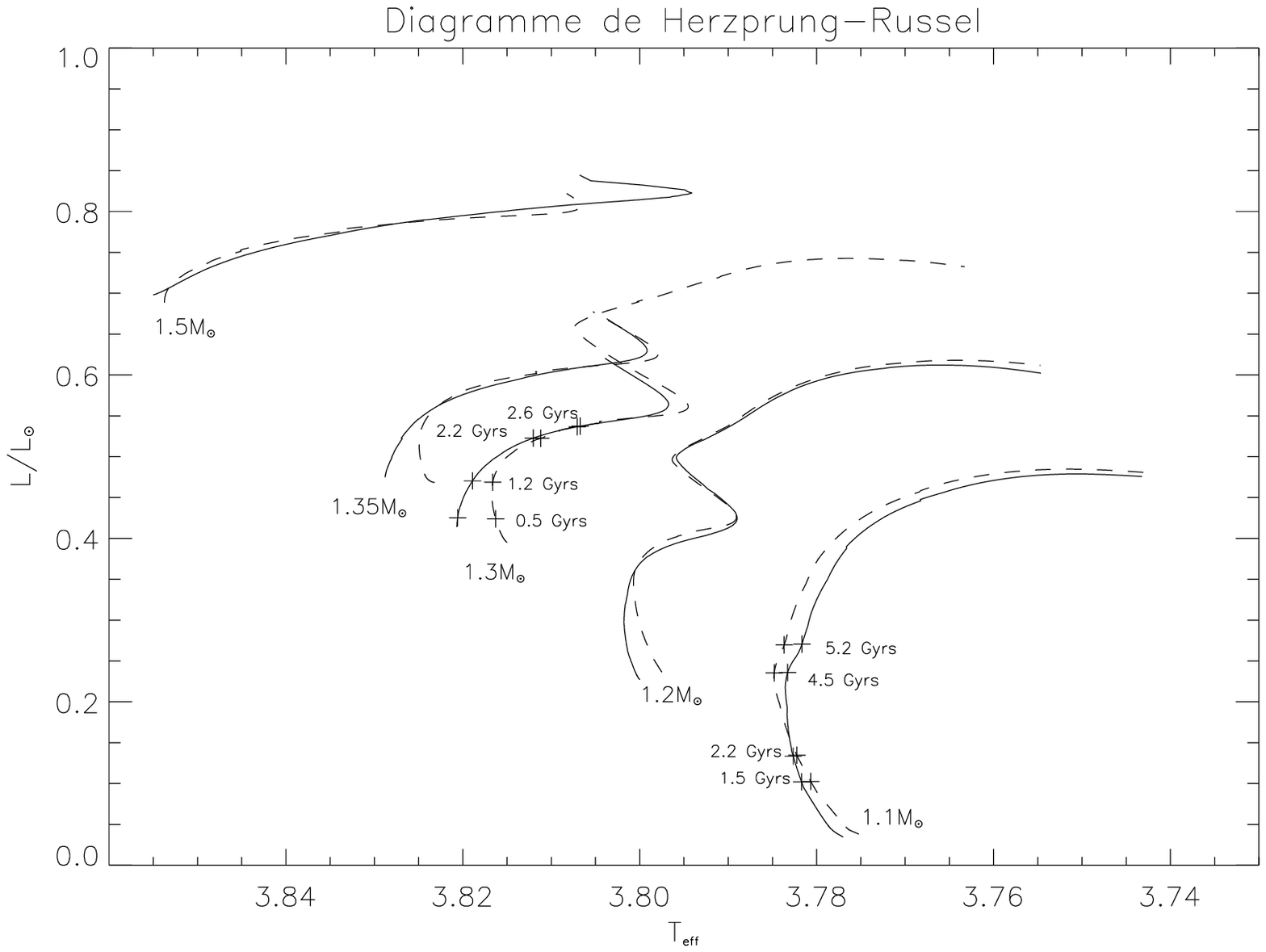}}
  \caption{Hertzprung-Russel diagram of evolution of models with (solid lines) and without (dashed lines) diffusion for different masses}\label{fig:HR}
\end{figure}

\begin{figure}
 \centering
 \resizebox{6.5cm}{4cm}{\includegraphics{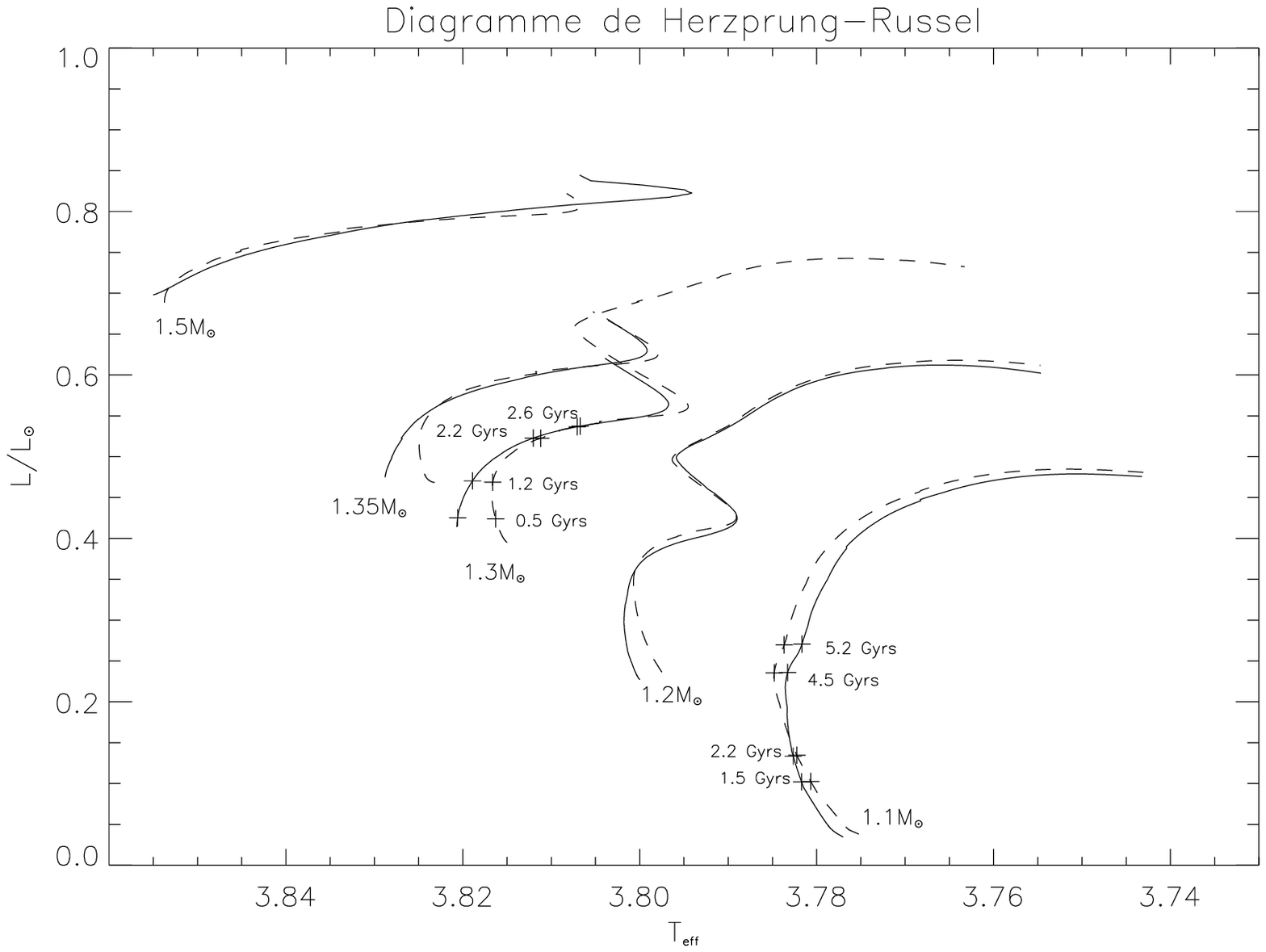}}
  \caption{Helium profile (a) and gradient of the sound velocity (b) in the models with (solid lines) and without (dashed line) diffusion for 1.3 $M_{\odot}$ at the crossing point of both tracks in the HR diagram, as a function of the acoustic depth.}\label{fig:gradhelium}
\end{figure}

Figure \ref{fig:gradhelium} displays the helium profiles and the gradients of the sound velocity in the models with (solid lines) and without (dashed lines) diffusion for 1.3 $M_{\odot}$ at the crossing point of the two tracks in the HR diagram, as a function of the acoustic depth, i.e. the time needed for the acoustic waves to travel between the surface and the considered region. In the former the helium gradient due to the diffusion is clearly seen. In the latter, the features around 700 s are due to the helium ionisation zones and the bottom of the convective zone lead to characteristic features at t = 2400 s. 

The partial reflections of the pressure waves on the regions of rapid variations in the sound velocity, such as the base of the convective zone or the helium ionisation zone, modulate the computed frequencies of the stellar spectrum. In the Fourier transform of the ``second differences'' (\cite{gough90}) of the different models, the position of the peaks correspond to the modulation periods, which are equal to twice the acoustic depth.

\begin{figure}
 \centering
 \resizebox{2.1cm}{2.1cm}{\includegraphics{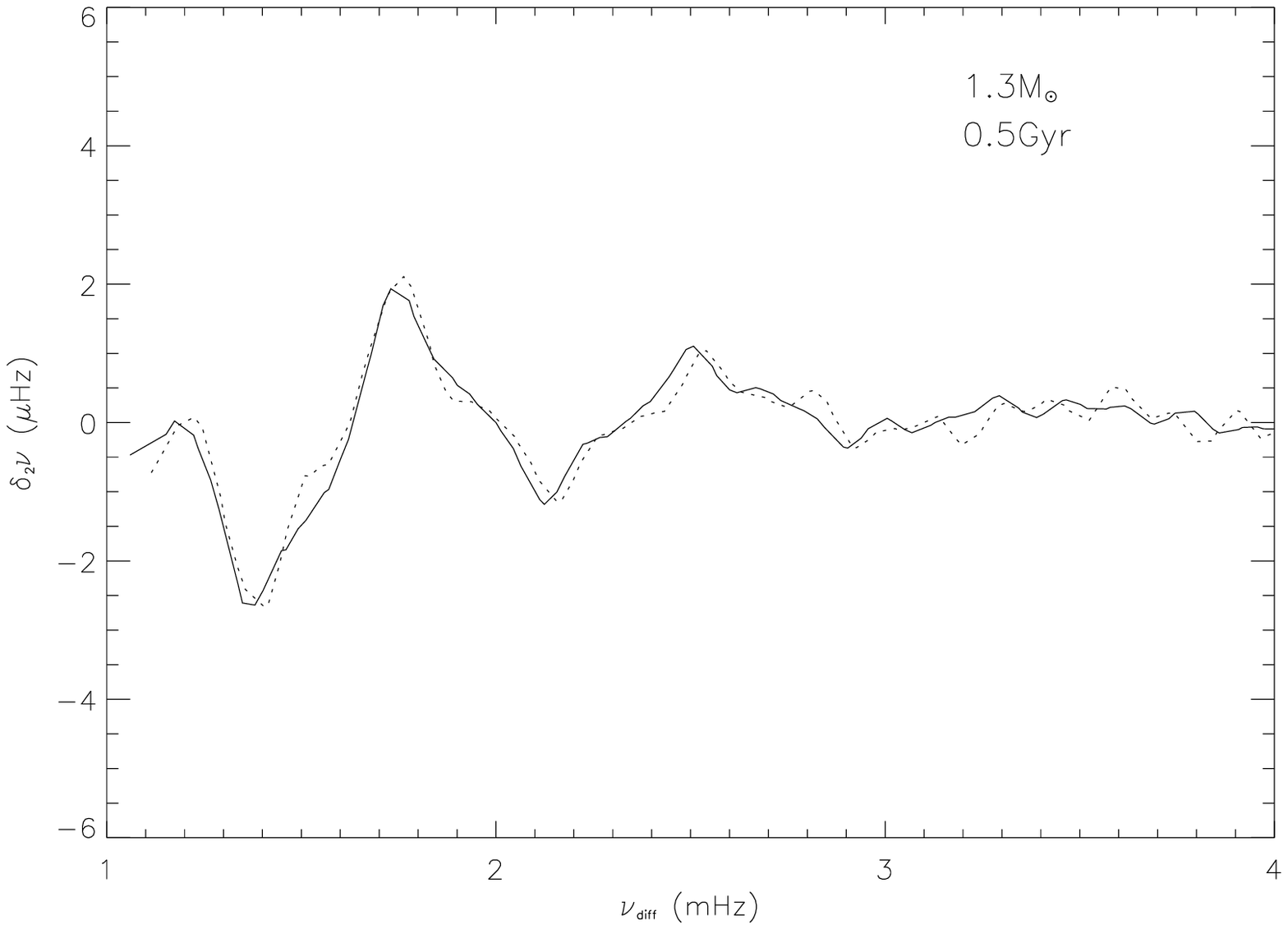}}
 \resizebox{2.1cm}{2.1cm}{\includegraphics{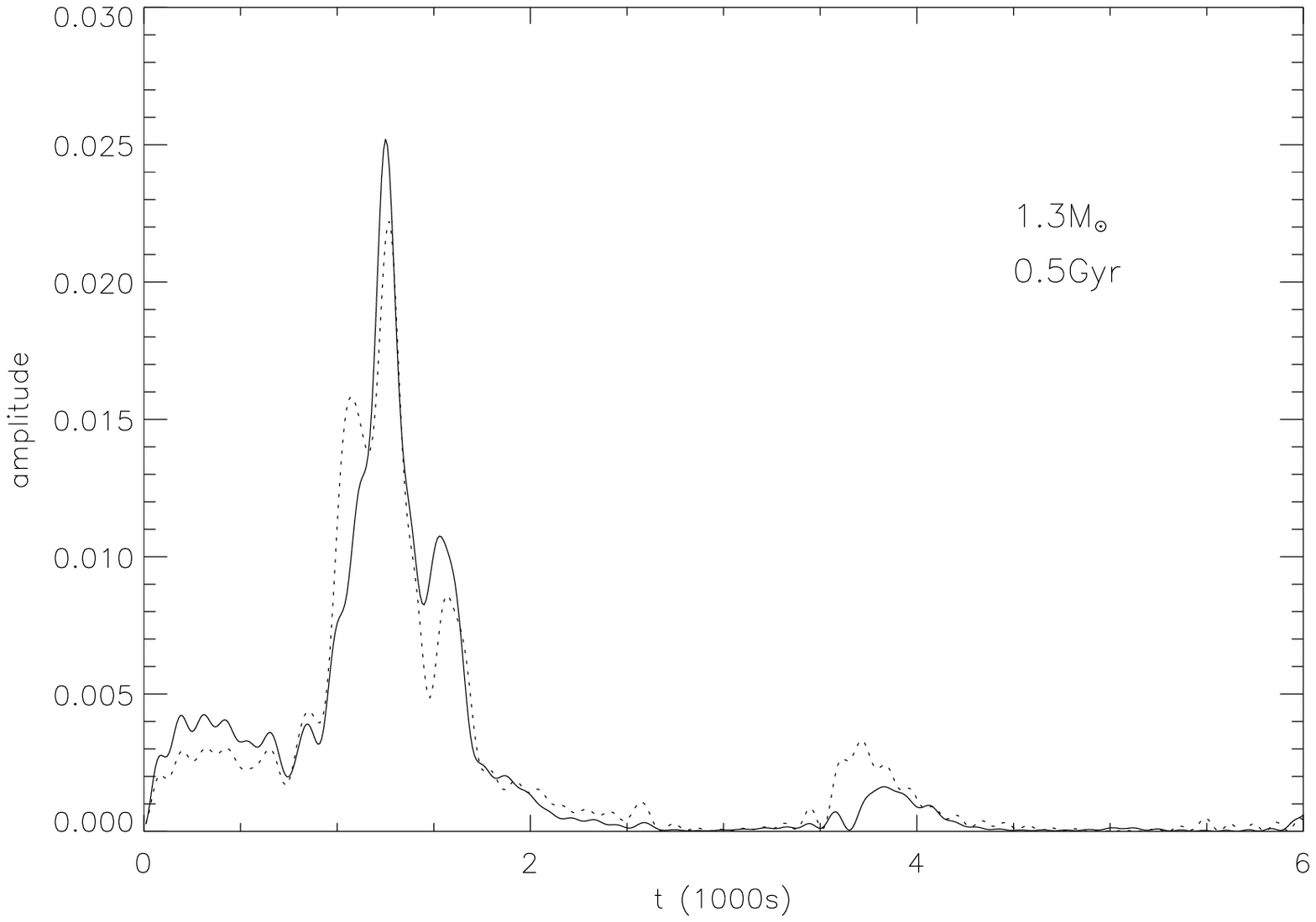}}
 \resizebox{2.1cm}{2.1cm}{\includegraphics{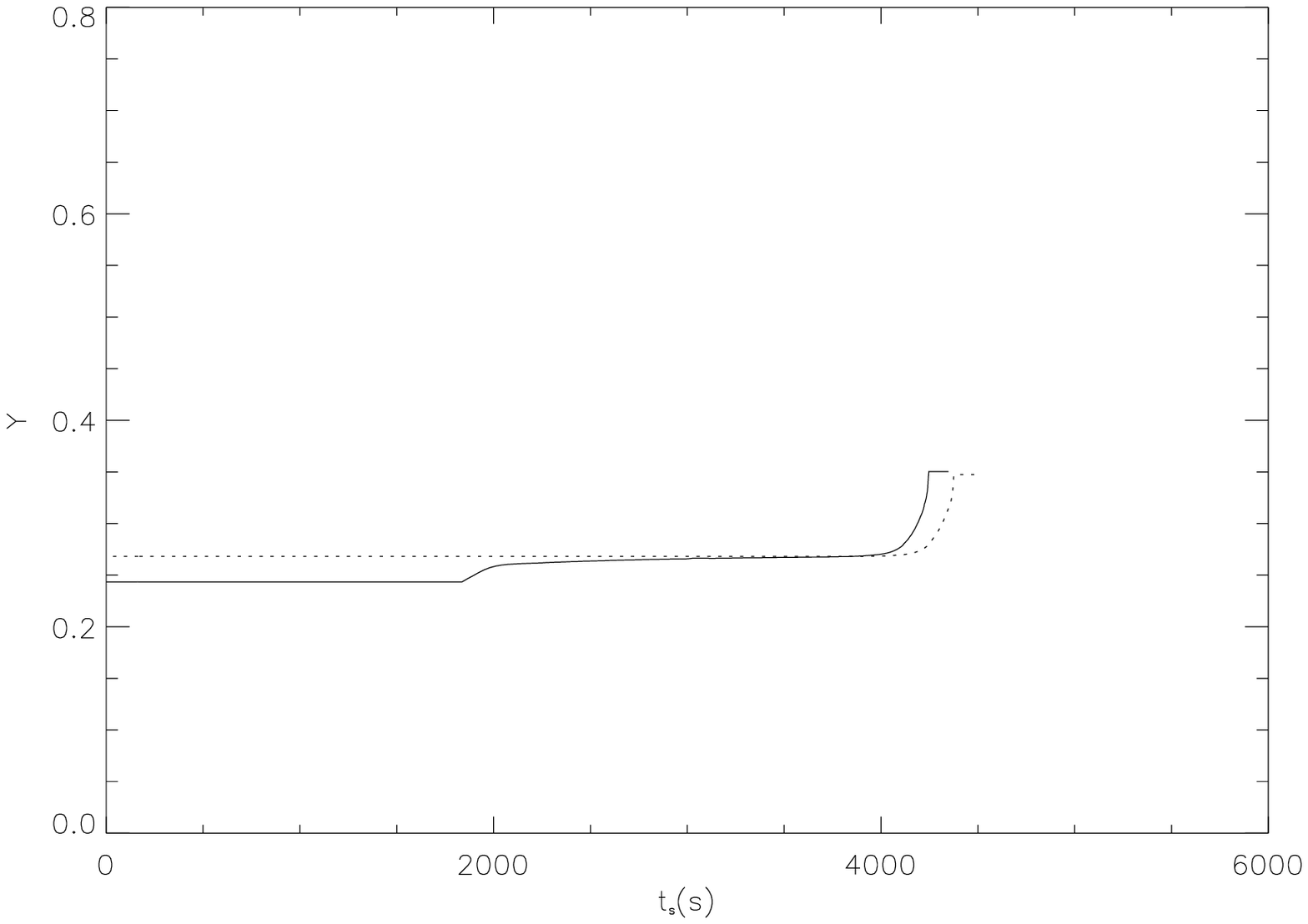}}
 \resizebox{2.1cm}{2.1cm}{\includegraphics{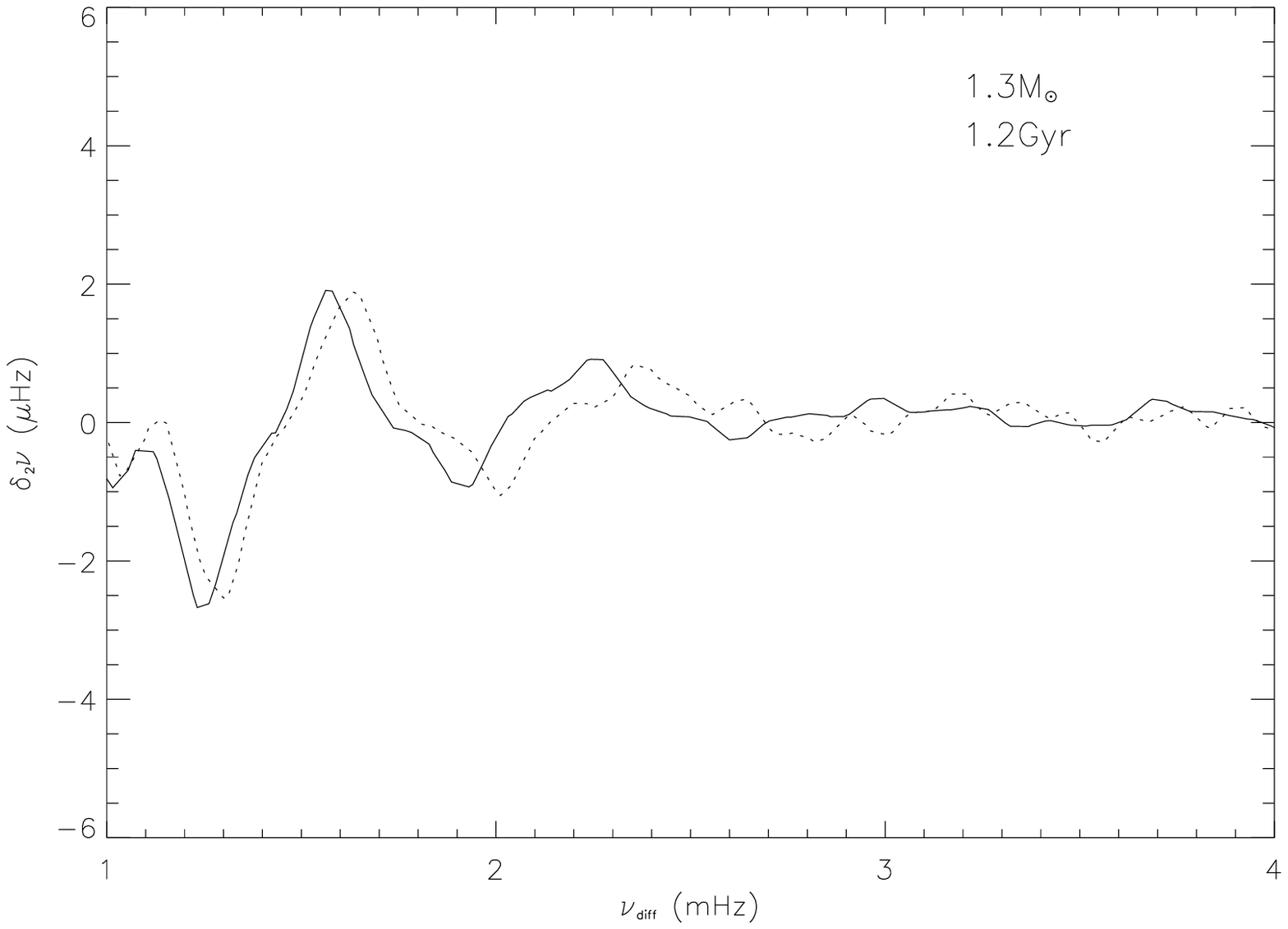}}
 \resizebox{2.1cm}{2.1cm}{\includegraphics{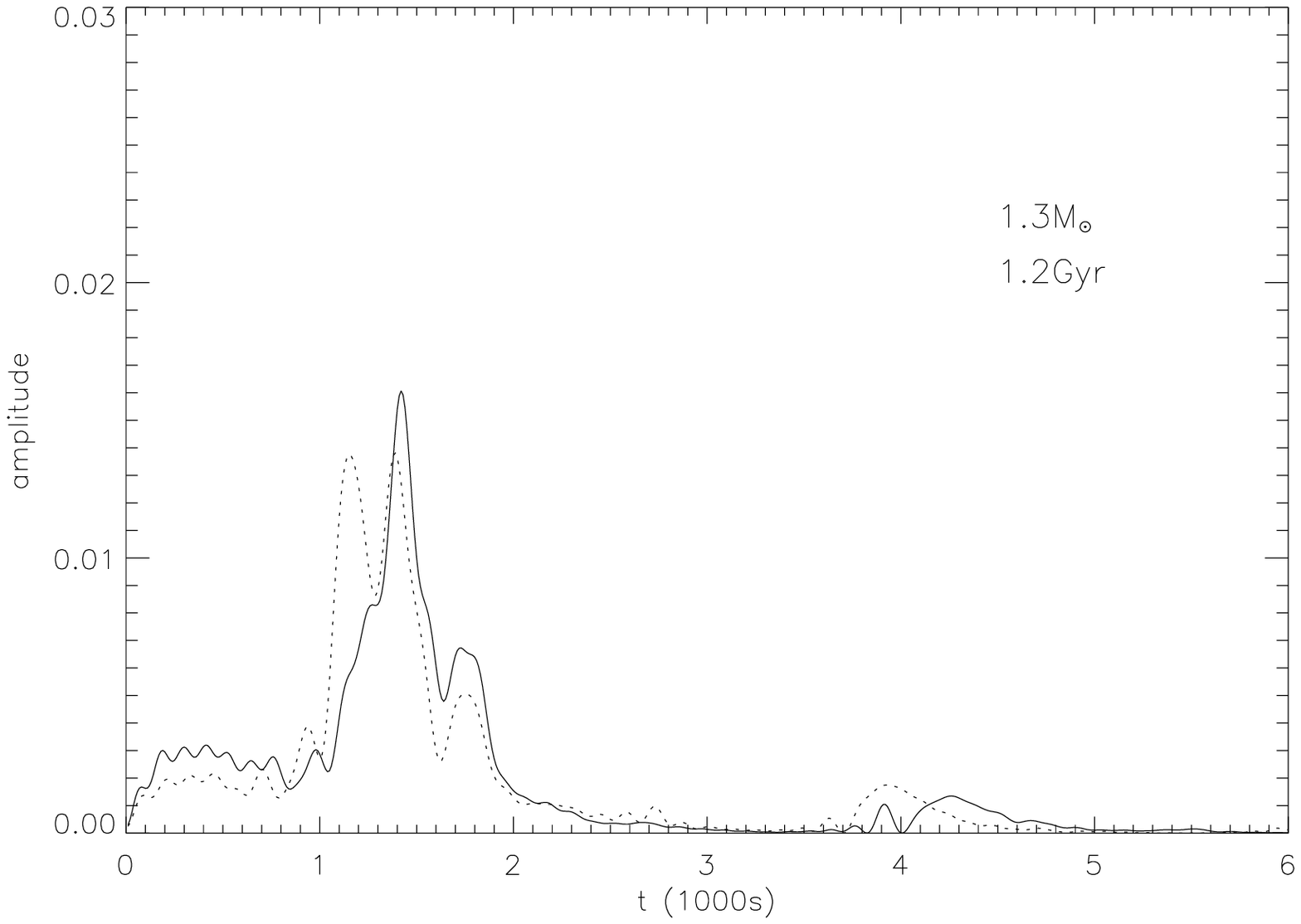}}
 \resizebox{2.1cm}{2.1cm}{\includegraphics{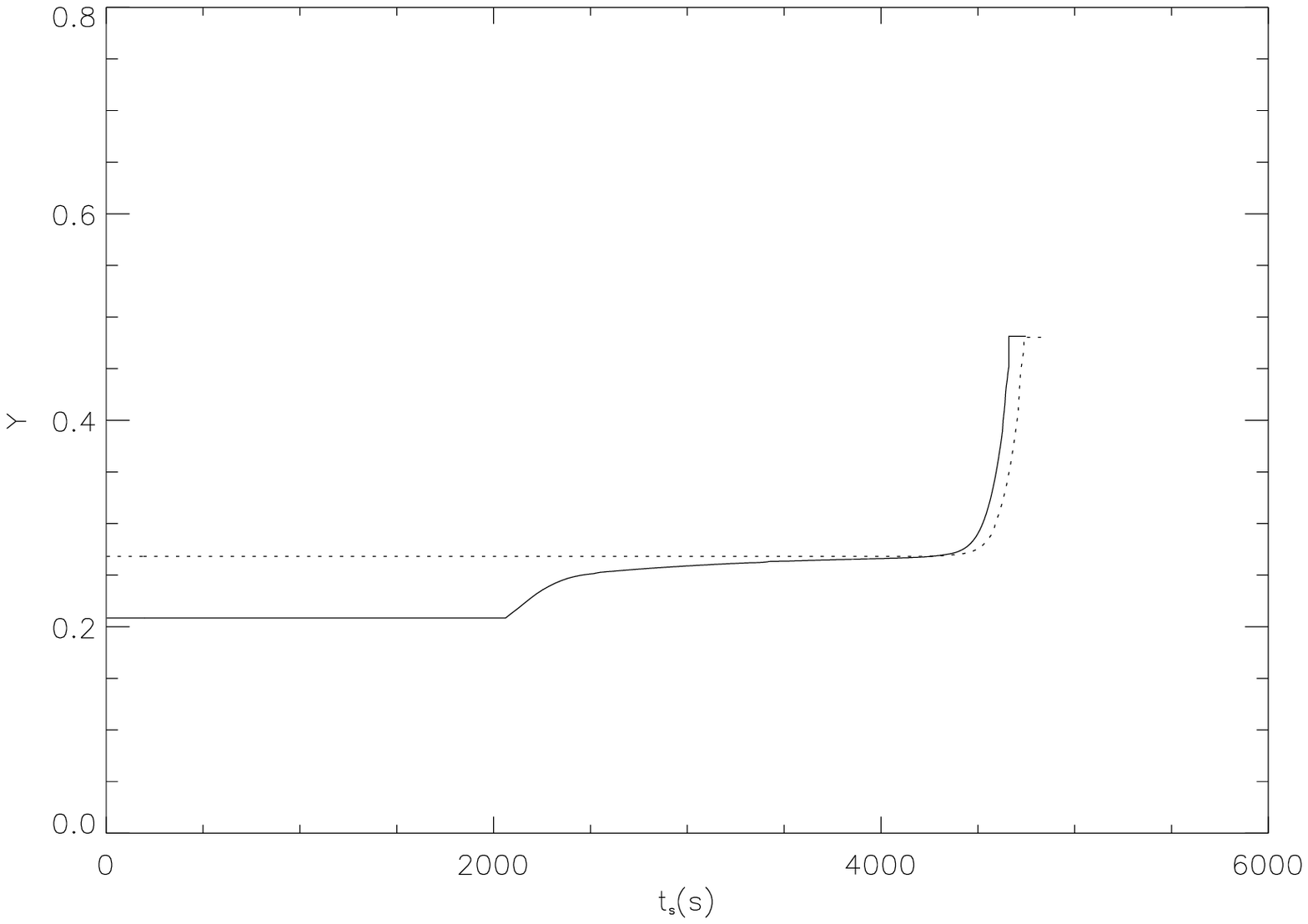}}
 \resizebox{2.1cm}{2.1cm}{\includegraphics{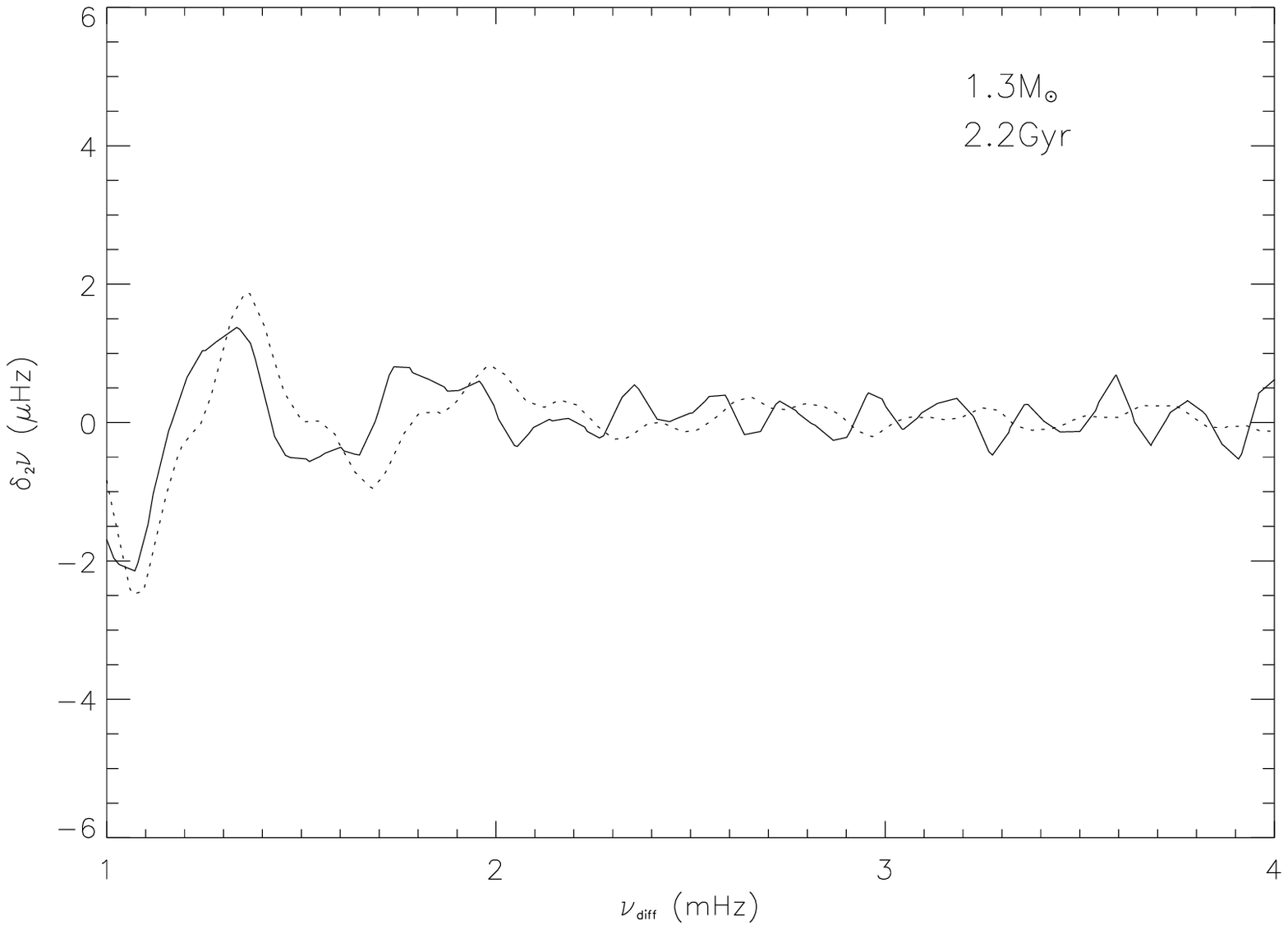}}
 \resizebox{2.1cm}{2.1cm}{\includegraphics{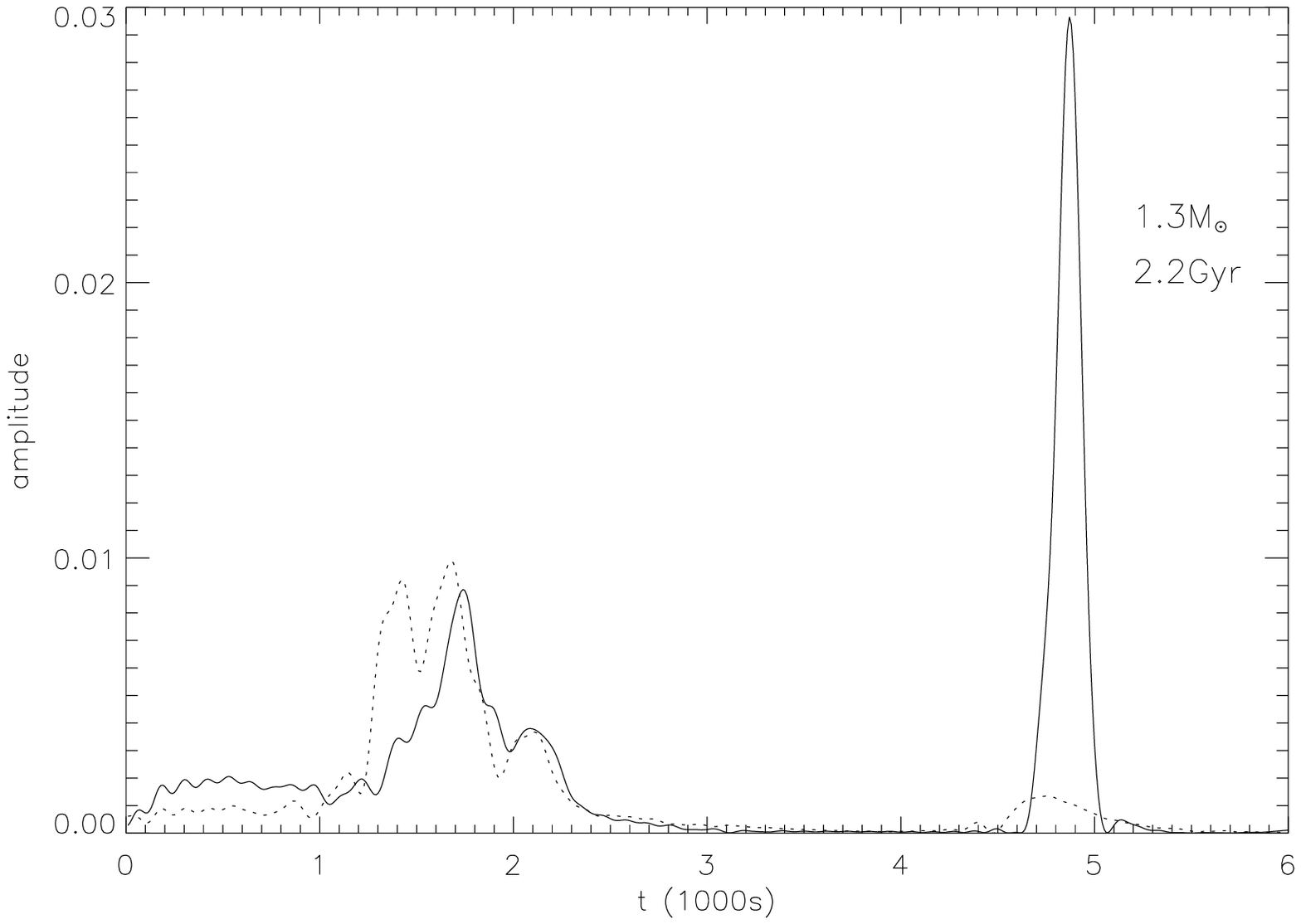}}
 \resizebox{2.1cm}{2.1cm}{\includegraphics{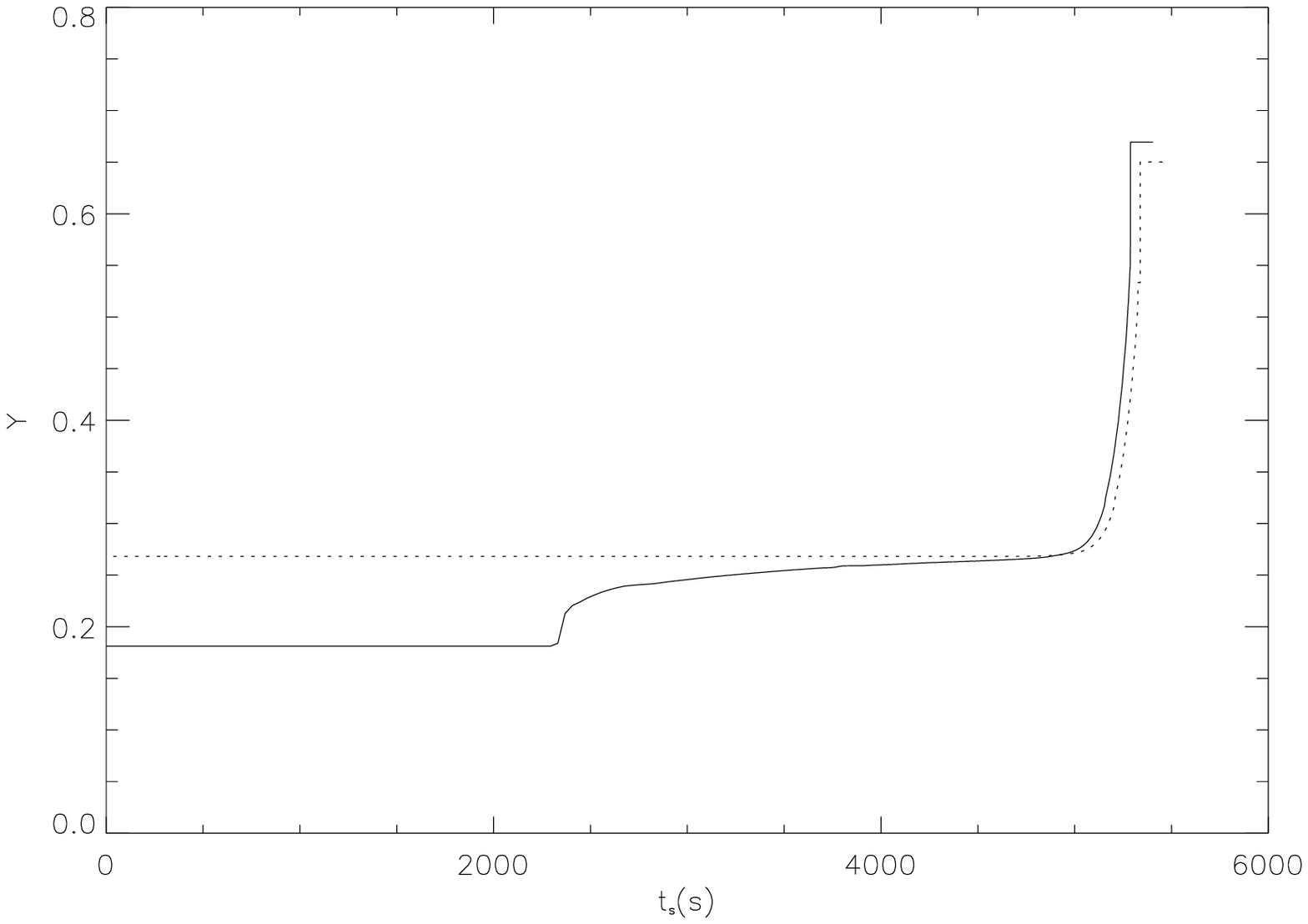}}
 \resizebox{2.1cm}{2.1cm}{\includegraphics{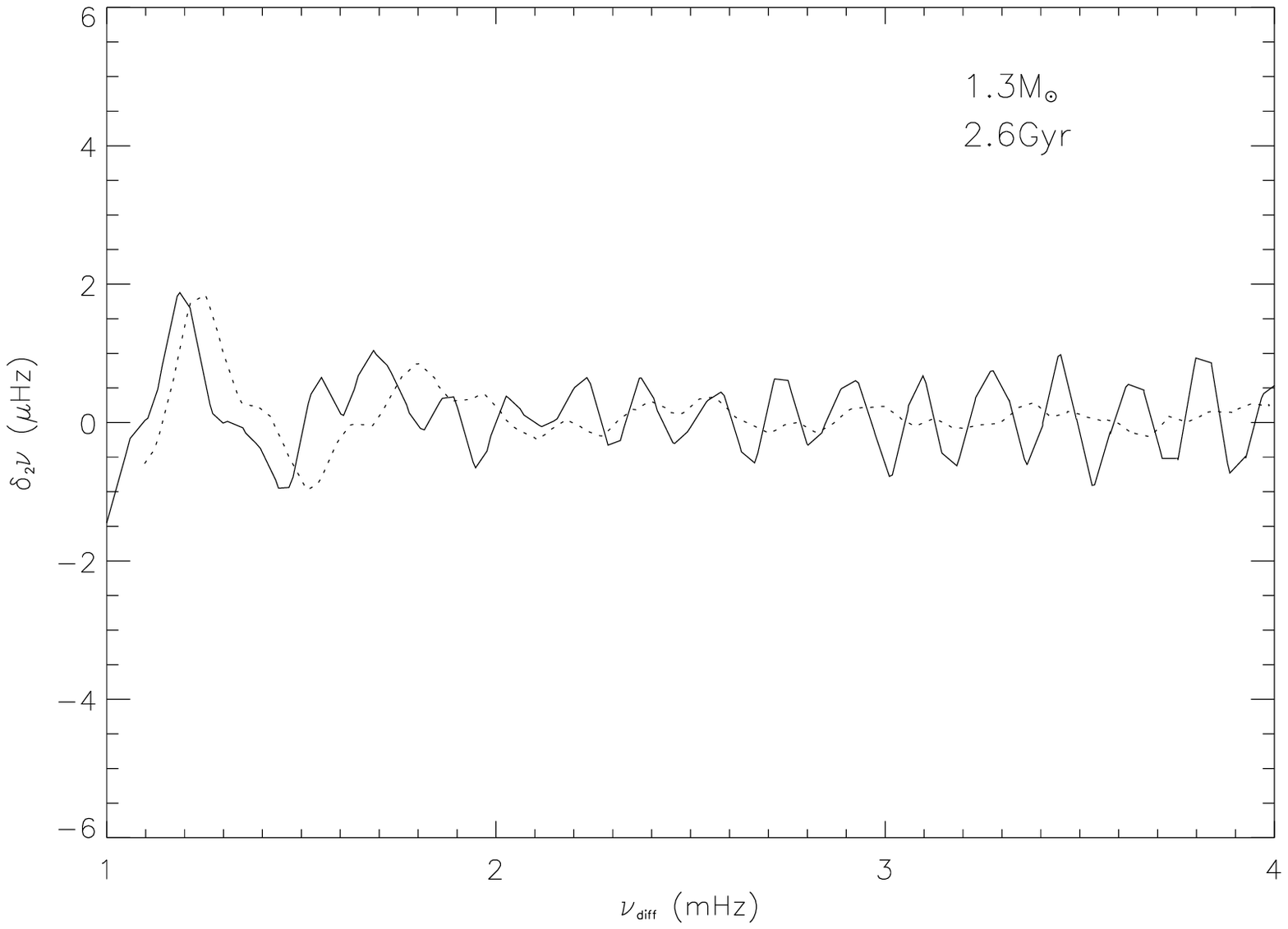}}
 \resizebox{2.1cm}{2.1cm}{\includegraphics{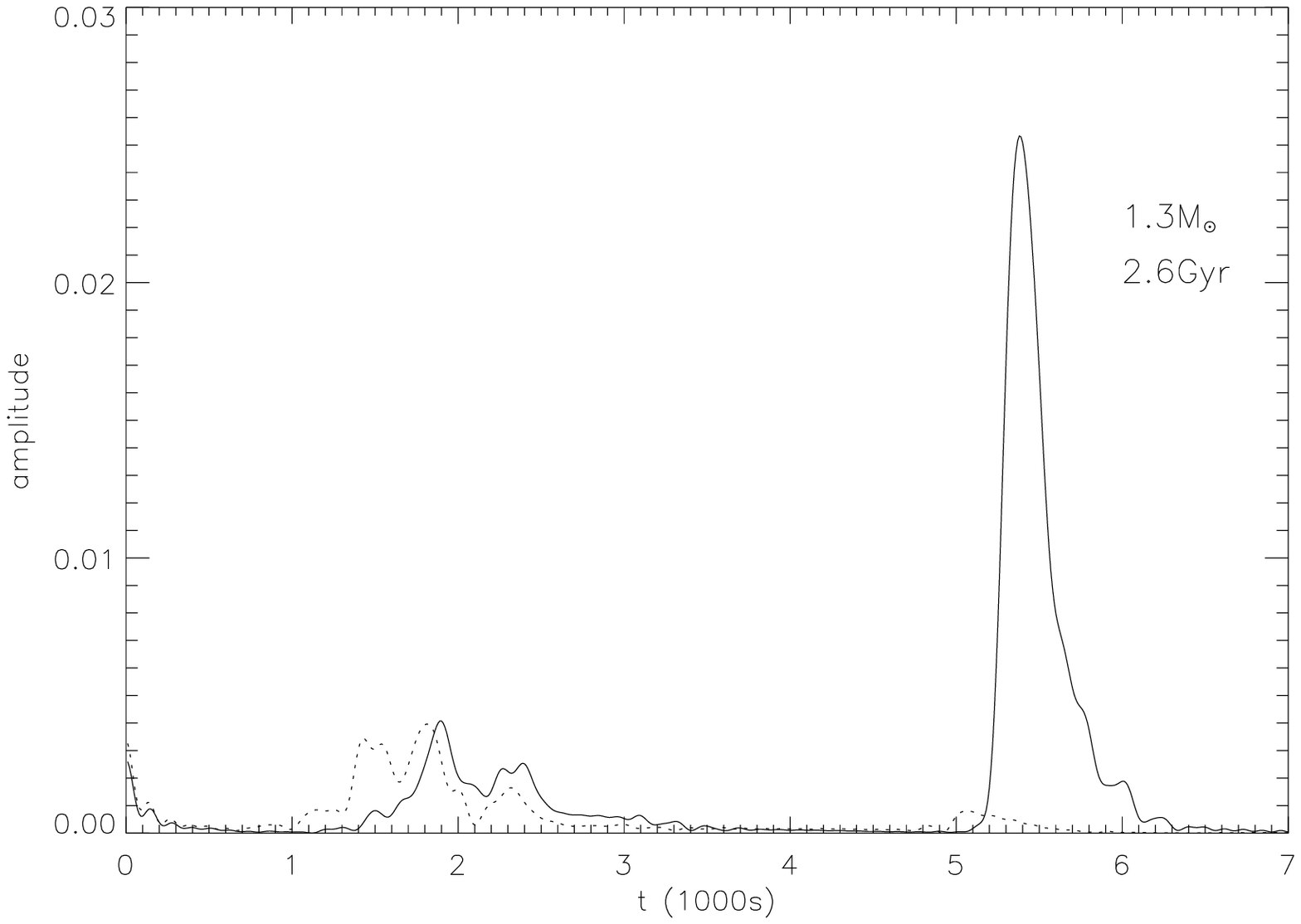}}
 \resizebox{2.1cm}{2.1cm}{\includegraphics{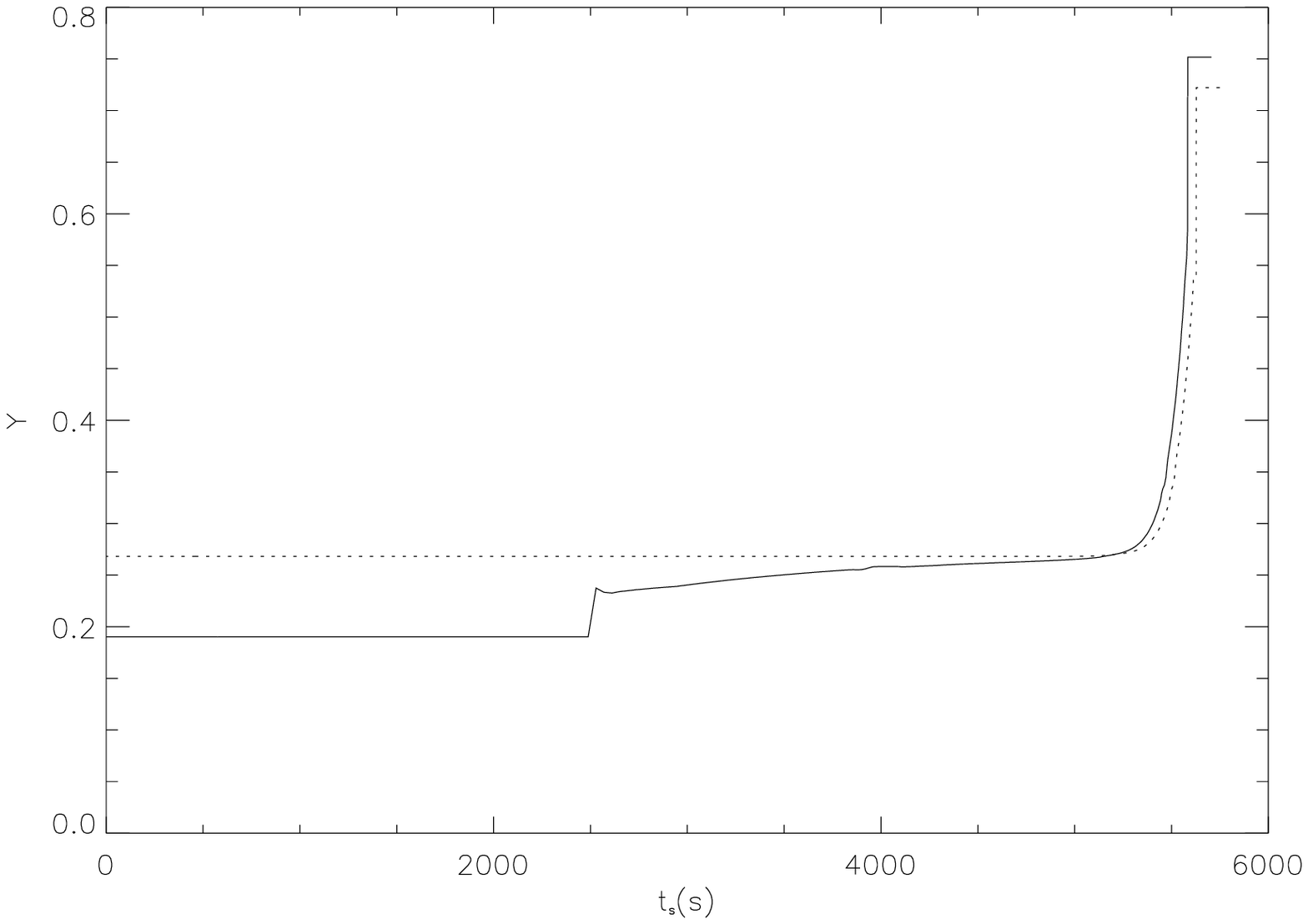}}
  \caption{Second differences (first column), Fourier transform of the second differences (second column) and helium profile (third column) of 1.3 $M_{\odot}$ at 0.5 Gyrs, 1.2 Gyrs, 2.2 Gyrs and 2.6 Gyrs.}\label{fig:sdiff13}  
\end{figure}

We present here results for models of 1.3 $M_{\odot}$. Other masses will be studied in a future paper (\cite{castro06}).
In Figure \ref{fig:sdiff13}, the peak due to the helium ionisation zone decreases in amplitude during the evolution, because of the decrease in helium concentration in the convective zone due to the diffusion. The peak due to the base of the convective zone undergoes a strong increase of amplitude due to the strong reflection of the pressure waves on the steep helium gradient at the moment when the convective zone sinks. Then, helium dilutes in the convective zone and the amplitude of the peak decreases. \\

 We studied the effect of helium diffusion on the second differences, which clearly show a gradient below the base of the convective zone. In future works, it would be interesting to introduce the radiative accelerations on metals, which can lead to metal accumulation in specific stellar layers, in the models and to take into account the different mixing processes.    

\bibliographystyle{aa}

\end{document}